\documentstyle[12pt,epsfig]{article}
\newcommand{\beeq}{\begin{equation}}
\newcommand{\ene}{\end{equation}}
\newcommand{\bea}{\begin{eqnarray}}
\newcommand{\ena}{\end{eqnarray}}
\newcommand{\no}{\noindent}

\newcommand{\T}{{\cal T}_k}

\newcommand{\nb}{\nonumber}

\begin{document}

\begin{titlepage} 
\begin{flushright}
 IFUP-TH 69/96\\
  Novenber  1996\\
\end{flushright}

\vspace{5mm}

\begin{center}

{\Large \bf Three-manifold invariants and 

\vskip 0.5truecm
their relation with the fundamental group}

\vspace{12mm}
{\large E.~Guadagnini ${}^{a,b}$ and  L.~Pilo ${}^{b,c}$}

\vspace{6mm}
(a) Dipartimento di Fisica dell'Universit\`a di Pisa, \\ 
Piazza Torricelli, 2. $\quad$ 56100 Pisa. Italy

\vspace{3mm}
(b) Istituto Nazionale di Fisica Nucleare, Pisa Italy. 
\vspace{3mm}

\vspace{3mm}
(c) Scuola Normale Superiore \\ 
Piazza dei Cavalieri 7. $\quad$ 56100 Pisa, Italy.

\vspace{6mm}
E-Mail: guadagni@ipifidpt.difi.unipi.it, $\quad$ pilo@ibmth.difi.unipi.it

\vspace{3mm}

\end{center}
\vspace{8mm}

\begin{quote}
\hspace*{5mm} {\bf Abstract.} 
 We consider the 3-manifold invariant $\, I(M) \, $ which is defined  by means of the
Chern-Simons quantum field theory and which coincides with the Reshetikhin-Turaev invariant. 
We present some arguments and numerical results supporting the conjecture
that, for nonvanishing  $\, I(M) \, $, the absolute value $\, | \,  I(M) \, | \, $ only depends on
the fundamental group $\pi_1 (M) \, $ of the manifold $ \, M \, $. For lens spaces, the conjecture
is proved when the gauge group is $\, SU(2) \, $. In the case in which the
gauge group is $\, SU(3) \, $, we present numerical computations confirming the conjecture.
\end{quote}
\end{titlepage}
\clearpage

\section{\bf Introduction}

Recently, new 3-manifold invariants \cite{wit,res}   have  been discovered; the algebraic aspects of
their  construction, which is based on the structure of simple Lie groups, are well understood
\cite{tur,km,lick1,mor,kau,koh,gua3}. However, the topological meaning of these invariants is still
unclear. Let us denote by $\, I(M) \, $ the invariant of the 3-manifold $ \, M \, $ which is
closed, connected and orientable; $\, I(M) \, $ is the invariant defined  by means of the
Chern-Simons quantum field theory \cite{wit,gua3} and coincides with the Reshetikhin-Turaev invariant
\cite{res,tur}.  In general, it is not known  how $\, I(M) \, $  is related, for
instance, to the homotopy class of $\, M \, $ or to the fundamental group of $\, M \, $.  In this
article we shall formulate the following

\begin{itemize} \item[]  ${\underline {\rm Conjecture}}$: for nonvanishing  $\, I(M) \, $, the
absolute value $\, | \,  I(M) \, | \, $ only depends on the fundamental group $\pi_1 (M) \, $. 
\end{itemize}

\noindent In the absence of a general proof, we shall verify the validity of the conjecture for
a particular class of manifolds: the lens spaces.  There are examples of lens spaces  $\, M_1 \, $
and $\, M_2 \, $ with the same fundamental group $\, \pi_1(M_1)
\simeq \pi_1 (M_2)\, $ which are not homeomorphic;  for all these manifolds, we shall prove
that  (for nonvanishing invariants) $\, | \,  I(M_1) \, | = | \,  I(M_2) \, |\, $  when
$\,   I(M) \,  $ is the invariant associated with the group $\, SU(2)\, $.  In the case in which the
gauge group is $\, SU(3) \, $, we will present numerical computations confirming the conjecture. Our
results are in agreement with the computer calculations for $\, SU(2) \, $ of Freed and Gompf
\cite{fregomp} and the expression of the $ \, SU(2) \, $ invariant obtained by Jeffrey \cite{jef}. 
Differently from
\cite{fregomp} and \cite{jef}, our approach is based exclusively on the properties of 
3-dimensional Chern-Simons quantum field theory. We shall use general surgery rules to compute
$\,   I(M) \,  $  and, in our construction,  invariance under Kirby moves is
manifestly satisfied.   

Our notations and conventions are described in section~2. The expression of the 
invariant $\, I(M) \, $ for a generic lens space is derived in section~3 and, for the 
gauge group $ \, SU(2) \, $, the validity of our conjecture is proved in section~4. The numerical
computations for the group $\, SU(3) \, $ are reported in section~5 and the conclusions are
contained in section~6.

\section{\bf Surgery rules}

The basic ingredient in the construction of the 3-manifold invariant $\, I(M) \, $ is a
polynomial invariant $\, E({\cal L}) \, $ for oriented, framed and coloured links $ \,
\{ \, {\cal L} \, \} \, \subset S^3 \, $. In the Chern-Simons field theory, this link invariant is
defined by the expectation values of the Wilson line operators \cite{wit}; each link component is
framed and its colour is given by an irreducible representation of a simple compact Lie group
which is called the gauge group. For example, when the gauge group is $ \, SU(N) \, $ and each
link component has colour corresponding to the fundamental representation of  $ \, SU(N) \, $,
$\, E({\cal L}) \, $ is determined by the skein relation \cite{wit,gua1} 
\beeq
q^{1/(2N)}\, E({\cal L}_+) \, - \, q^{-1/(2N)}\, E({\cal L}_-) \; = \; ( \, q^{1/2} - q^{-1/2} \, )
\, E({\cal L}_0) \qquad , 
\label{skein}
\ene
where $\, q = \exp(-i 2 \pi/k) \,$ is the deformation parameter and $\, k\, $ is the renormalized
coupling constant of the Chern-Simons field theory. The standard skein-related links $ \, {\cal L}_+ \,
$,  $ \, {\cal L}_- \, $ and $ \, {\cal L}_0 \, $ correspond to a configuration with
over-crossing, under-crossing and no-crossing respectively. Moreover, under an elementary
$\, \pm 1 \, $ modification of the framing of a link component, $\, E({\cal L}) \, $ gets
multiplied by the factor  $ \, q^{\pm (N^2 -1)/2N} \, $. Finally, the factorization property
\cite{wit,gua1} which holds for the distant union of links fixes the normalization of the unknot 
with preferred framing 
\beeq
 E_0[ \, {\rm fund.}\, ]  \; = \; (\, q^{N/2} - q^{-N/2} \, ) / ( \, q^{1/2} - q^{-1/2} \, ) 
\qquad .  
\label{unknot}
\ene
In general, the colour which characterizes one link component is an element of the algebra $\,
{\cal T} \, $ which coincides with the complex extension of the representation ring of the gauge
group. The sum operation in this algebra extends by linearity to $\, E({\cal L}) \, $; whereas the
product operation in the colour algebra $\, {\cal T} \, $ simply corresponds to the satellites
obtained from the companion links by standard cabling \cite{mor,gua2}.  For unitary groups, the
fundamental skein relation (\ref{skein}), the normalization ({\ref{unknot}) of the unknot and the
correspondence between cabled components and higher-dimensional representations of the gauge
group uniquely determine the values of the link invariant  $\, E({\cal L}) \, $ for arbitrary
coloured link components. 

Let us denote by $\, {\cal L}_1 \# {\cal L}_2 [\rho ]\, $ the connected sum of the links $\,
{\cal L}_1 \, $ and  $\, {\cal L}_2 \, $ in which the component which connects these two links has
colour given by the irreducible representation $ \, \rho \, $ of the gauge group.  From the
properties of the Chern-Simons field theory it follows that \cite{wit,gua2}
\beeq
E( \, {\cal L}_1 \# {\cal L}_2 [\rho ]\, ) \; = \; { E({\cal L}_1) \, E({\cal L}_2) \over
E_0[ \, \rho \, ]}
\quad ,  
\label{somcon} 
\ene
where $\, E_0[ \, \rho \, ] \, $ is the value of the unknot with preferred framing and colour $
\, \rho \, $. 

For integer values of the Chern-Simons coupling constant $ \, k \, $ ($ \, k =1,2,3,... \, $),
the set of vanishing link invariants defines  an ideal $ \, {\cal I}_k \, $ of $\, {\cal T} \, $.
Thus, for fixed integer $ \, k \, $, the colour states belong to the algebra \cite{gua2} of the
equivalence classes 
\beeq
{\cal T}_{(k)} \; = \; {\cal T} \; / \; {\cal I}_k \qquad . 
\label{reduced}
\ene
Usually, $\, {\cal T}_{(k)} \, $ is of finite order \cite{gp3} and, for appropriate values of $ \, k
\, $,   $\, {\cal T}_{(k)} \, $ is isomorphic with the Verlinde algebra \cite{verl} which
is determined by of the fusion rules  of certain two-dimensional conformal models \cite{wit}. 
We shall now concentrate on $\, {\cal T}_{(k)} \, $ when the gauge group $ \, G \, $ is $\,
SU(2) \, $ \cite{gua2} or $\, SU(3) \, $ \cite{gp1}.    For $ \, G = SU(2) \, $ and $ \, k=1\, $, 
$\, {\cal T}_{(1)} \, $ is isomorphic with the group algebra of $ \, Z_2 \, $, which is the center of
$ \, SU(2) \, $. For $ \, G = SU(2) \, $ and $ \, k\geq 2\, $, the ideal $ \, {\cal I}_k \, $ is
generated by the representation with $ \, J= (k-1)/2 \, $ and $\, {\cal T}_{(k)} \, $ is of order
$\, (k-1) \, $. For $ \, G = SU(3) \, $ and $ \, k=1,2\, $, the algebra (\ref{reduced}) is
isomorphic with the group algebra of $ \, Z_3 \, $, which is the center of $\, SU(3) \, $. For $
\, G = SU(3) \, $ and $ \, k\geq 3\, $, the ideal $ \, {\cal I}_k \, $ is generated by the two
irreducible representations with  Dynkin labels $ \, (k-1,0) \, $ and $(k-2,0 ) \, $; in this
case,  $\, {\cal T}_{(k)} \, $  is of order $\, (k-1)(k-2)/2 \, $.

We shall denote by  $\, \{ \, \psi[\, i \,] \, \}\, $ (with $ \, i =1,2, ...,  {\rm dim}( {\cal
T}_{(k)} ) $) the elements of a basis in $\, {\cal T}_{(k)} \, $. When $\, G =SU(2) \, $ and $
\, k \geq 2\, $ or when $\, G =SU(3) \, $ and $ \, k \geq 3\, $,  each $\, \psi[\, i\, ]\, $
represents  the equivalence class of an irreducible representation of the gauge group. For low
values of $ \, k \, $, $\, \psi[\, i\, ]\, $ corresponds to an irreducible representation of the
gauge group up to a nontrivial multiplicative factor \cite{gua2,gp1}.     The unit in $\, {\cal
T}_{(k)}\, $ will be denoted by $\, \psi[1]\, $; $\, \psi[1]\, $ is the class defined by the 
trivial representation. 

Let us now consider the definition of the 3-manifold invariant $ \, I(M) \, $.  Each
3-manifold $\, M \, $, which is closed, connected and orientable, admits a  surgery
presentation \cite{rol} given by Dehn surgery on $\, S^3 \, $. Each ``honest" \cite{rol} surgery
instruction can be represented by a framed link $\, {\cal L} \subset S^3\, $ with components $\{
\, {\cal L}_b \, \}$ with $ \, b = 1,2,...  \, $. The surgery link $\, {\cal L} \, $ is not
oriented and an integer surgery coefficient $\, r_b\, $ is attached to the component $\, {\cal
L}_b \, $. The framing $\, {\cal L}_{bf}\, $ of $\, {\cal L}_b \,  $ is specified by the linking
number 
\beeq
\ell {\rm k} \,  ( {\cal L}_b , {\cal L}_{bf} ) \; = \; r_b \quad . 
\label{eq:fra}
\ene 
The surgery link associated to the manifold $ \, M \, $ is not unique. Indeed, if 
the surgery links $\, \cal L \, $ and $\, {\cal L}^\prime \, $ are related by
a finite sequence of Kirby moves, the corresponding manifolds are homeomorphic \cite{kir}. 
Therefore,  each 3-manifold $\, M\, $ is characterized by a class of ``equivalent" surgery links 
in $\, S^3\, $,  where ``equivalent" links means links related by Kirby moves. 

Let $\, {\cal L} \subset S^3\, $ be a surgery link for the manifold $ \, M \, $.  The  invariant
$\, I(M) \, $  is defined in terms of the expectation value $ \, E( {\cal L}  ) \, $ of the
Wilson line operators associated with the surgery link $\, {\cal L} \, $. More precisely,   
one introduces an (arbitrary) orientation and  a particular colour state $\, \Psi_0 \, $ for each
component of $\, {\cal L} \, $.  For fixed integer $ \, k \, $, the surgery colour state   $\,
\Psi_0 \in {\cal T}_{(k)}\, $ is 
\cite{res} 
\beeq
\Psi_0 \; = \; a_k \, \sum_i \, E_0[\, i\, ] \, \psi [\, i\, ] \qquad ,    
\ene
where the sum is performed over all the elements $\{ \,  \psi[\, i\, ]\,  \}$ of the basis of
$\, {\cal T}_{(k)}\, $. The coefficients $\, \{ \, E_0[\, i\, ] \, \} \, $ coincide with the
expectation values of the unknot with preferred framing and colour $ \,  \psi[\, i\, ]\,  $.  When
the gauge group $\, G \, $ is $\, SU(2)\, $, $\, a_k \, $ is given by \cite{gua2}
\beeq
a_k \, = \, \left \{ \begin{array}{cc} 1/\sqrt{2} & \quad k \, = \, 1 \quad \null \\
\sqrt{\frac{2}{k}} \sin{(\pi /k)} & \quad k \, \geq 2 \quad ;
\end{array} \right.
\ene
 whereas, when $\, G = SU(3)\, $, one has \cite{gp1}
\beeq
a_k \, = \, \left \{ \begin{array}{cc} 1/\sqrt{3} & \quad k \, = \, 1,2 \\
16 \cos(\pi/k) \sin^3(\pi/k)/ (k \sqrt{3}) & \quad k\, \geq 3 \quad .
\end{array} \right.
\ene
We shall  denote by $\, \sigma ({\cal L})\, $ the signature of the linking matrix associated
with $\, \cal L \, $;  $\, \sigma ({\cal L})\, $ does not depend on the choice of the orientation
of $ \, \cal L \, $.  Let us define the function  $\, I({\cal L})\, $  by means of the
relation \cite{res} 
\beeq
I ({\cal L}) \; = \;  \exp \left [\, i\, \theta_k \, \sigma ({\cal L})\,
\right ] \; \,  E ({\cal L})  \qquad , \label{inv} 
\ene
where, for $ \, G=SU(2) \, $, the phase factor $ \, e^{i \theta_k } \, $ is \cite{res,gua2}    
\beeq 
e^{i \theta_k} \, = \, \left\{ \begin{array}{cc} \exp \left (- i \pi /4 \right ) & \quad k = 1
\quad
\null
\\ \exp \left [i \pi 3(k-2)/(4k) \right ] & \quad k \geq 2 \quad ; \end{array} \right.
\ene
and, for $\, G = SU(3) \, $, the phase factor is \cite{gp1}  
\beeq 
e^{i \theta_k} \, = \, \left\{ \begin{array}{cc} \exp \left ( i \pi /2 \right ) & \quad k = 1
\quad
\null \\ \exp \left (- i \pi /2 \right ) & \quad k = 2 \quad \null \\
\exp \left (-i 6 \pi /k \right ) & \quad k \geq 3 \quad . \end{array} \right.
\ene
It can be verified \cite{res,tur,gua2,gp1} that $\, I ({\cal L}) \,$ is invariant under Kirby moves 
and then it represents a topological invariant for the 3-manifold $\, M \, $. In what follows, we
shall denote this invariant by $ \, I(M) \, $. 

It should be noted that the  multiplicative phase factor in (\ref{inv}) is not a matter of
convention (or choice of  framing); the presence of the term $\, \exp \left [\, i\, \theta_k \,
\sigma ({\cal L})\, \right ] \, $ in (\ref{inv}) guarantees the invariance of $\, I ({\cal L}) \,$
under Kirby moves. According to the definition  (\ref{inv}), the normalization of the 3-manifold
invariant $ \, I(M) \, $ is fixed by $ \, I(S^3) =1\, $. 

In order to compare  $ \, I(M) \, $ with the expressions obtained in \cite{fregomp,jef}, we need
to produce the relation between the link invariants and the representation matrices of the
mapping class group of the torus.  

\begin{figure}[h]
\vskip 0.9 truecm 
\centerline{\epsfig{file=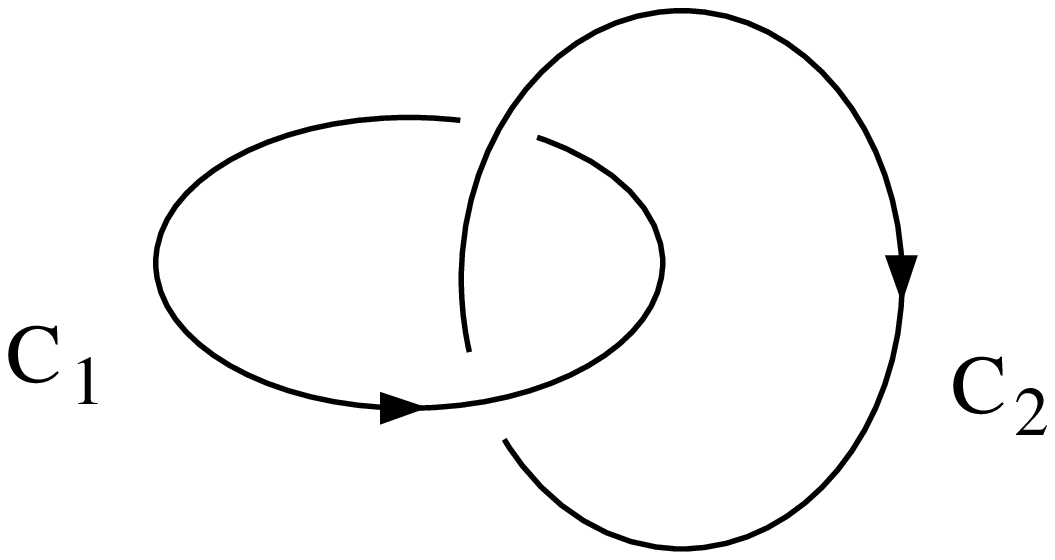,height=3cm,width=5cm}}
\vskip 0.9 truecm 
\centerline {{\bf Figure 1}}      
\vskip 0.9 truecm 
\end{figure}

\noindent Let us consider the Hopf link in $\, S^3\, $, shown in Figure~1; let the two link
components $\, C_1 \, $ and $\, C_2 \, $ have preferred framings and colours $ \, \psi [ \, i \, ]
\, $ and $ \, \psi [ \, j \, ] \, $ respectively.  The associated Chern-Simons expectation value 
is denoted by 
\beeq
H_{ij} \; =\; E (\, C_1, \, ,\, \psi[\, i\, ] \, ; \, C_2, \, , \, \psi[\, j\, ] \, ) \qquad . 
\ene
The complex numbers $ \, \{ \, H_{ij} \, \} \, $ where $ \, i,j =1,2, ...,  {\rm dim}( {\cal
T}_{(k)} ) $ can be understood as the matrix elements of the so-called Hopf matrix $\, H\, $.  
Note that  $\, H\, $ is symmetric  and that $ \, E_0[ \, i \, ] = H_{1i} = H_{i1}  \, $. 
Let $\, Q(i)\, $  be the value of the quadratic Casimir operator of the
irreducible representation of the gauge group which is associated with an element of the class
$\psi[\, i\, ]$.  One can show \cite{gp3} that the matrices   
\beeq
X_{ij} \; = \; a_k \, H_{ij} \qquad ; \qquad 
Y_{ij} \; = \; q^{Q(i)} \, \delta_{i j} \qquad  ; \qquad 
C_{ij} \; = \; \delta_{i j^\ast} \qquad . \label{eq:modu}
\ene
give  a projective representation of the modular group
\beeq
 X^2  \; = \; C \qquad;
\ene
\beeq
\left( \, X \> Y \, \right )^3 \; = \; e^{- \, i \, \theta_k} \> C \qquad .
\ene
This representation is isomorphic with the representation obtained in two-dimensional
conformal field theories \cite{wit};  $\, X \, $  corresponds to the $\, S\,  $ matrix
of the conformal models and $ \, Y \, $ is the analogue of the $ \, T \, $ matrix.

\section{\bf Lens Spaces}
Lens spaces, which are characterized by two integers  $ \, p \, $ and $ \, r \, $, will be
denoted by  $\, \{ \,  L_{p/r} \, \} \, $. The fundamental group of $\, L_{p/r}\, $ is $\, Z_p\,
$. Two lens spaces $\, L_{p/r} \, $ and $\, L_{p^\prime/r^\prime}\, $ are homeomorphic if and
only if $\, |p| = |p^\prime| \, $ and $\, r =  \pm r^\prime \; (\mbox{mod } p)\, $ or $\, r
r^\prime = \pm 1 \; (\mbox{mod } p)\, $. Thus, we only need \cite{rol} to consider the case in
which $ \, p >1\, $  and $ \, 0 < r < p \, $; moreover, $ \, r \, $ and $ \, p \, $ are 
relatively prime.  The lens spaces $\, L_{p/r} \, $ and $\, L_{p^\prime/r^\prime}\, $ are
homotopic if and only if $|p|= |p^\prime|$ and $r r^\prime = \pm $ quadratic residue $ (\mbox{mod
} p)\, $.  Consequently, one can find examples of lens spaces which are homotopic  but are not
homeomorphic; for instance,  $\, L_{13/2}\, $ and $\, L_{13/5}\, $. One can also find examples of
lens spaces which are not homeomorphic and are not homotopic but have the same fundamental group;
for instance,  $\, L_{13/2}\, $ and $\, L_{13/3}\, $.

One possible surgery instruction corresponding to  the lens space $\, L_{p/r} \, $ is given
the unknot \cite{rol} with rational surgery coefficient $\, (p/r) \, $. From this surgery
presentation one can derive \cite{rol} a ``honest" surgery presentation of $\, L_{p/r} \, $ by
using  a continued fraction decomposition of the ratio $\, (p/r) \,$ 
\beeq
\frac{p}{r} \, = \, z_d \, - \, \frac{1}{\; z_{d-1}\; -\;  \frac{1}{\; \ddots \; - \; 
\, \frac{1}{z_1}}} \qquad , 
\ene   
where $\, \{z_1, \, z_2, \cdots, \, z_d \} \,$ are integers. The new surgery link
$\, {\cal L} \, $ corresponding to  a ``honest" surgery presentation of $\, L_{p/r} \, $ is a
chain with $ \, d \, $ linked components, as shown in Figure~2, and the integers $\, \{z_1, \,
z_2, \cdots, \, z_d \} \,$ are precisely the surgery coefficients. 

\begin{figure}[h]
\vskip 0.9 truecm 
\centerline{\epsfig{file=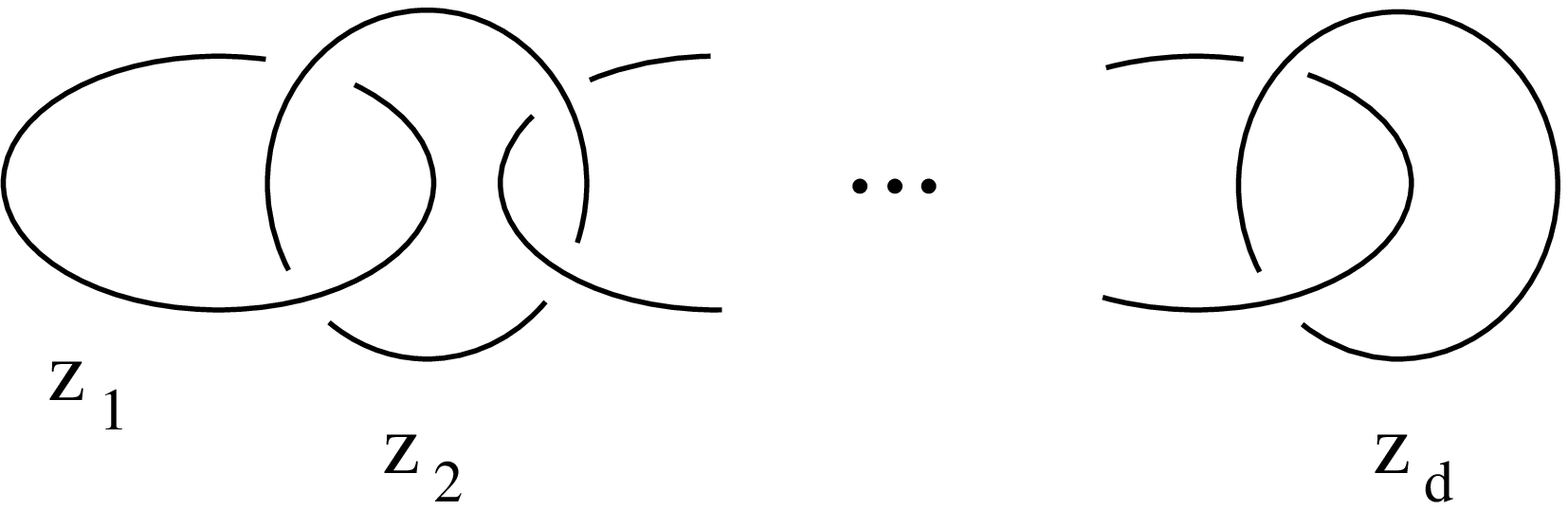,height=3cm,width=6cm}}              
\vskip 0.9 truecm 
\centerline {{\bf Figure 2}}      
\vskip 0.9 truecm 
\end{figure}

\noindent According to the definition (\ref{inv}), the lens space invariant is given by 

\bea
&&I(L_{p/r}) \; = \; e^{i \theta_k \sigma ({\cal L}) } \, ( \, a_k \, )^d \, \sum_{j_1, \cdots ,
j_d \in \T} \;  \prod_{i=1}^d \left ( q^{z_i \, Q(j_i)} \right ) \, \times \nb  \\
&&
\qquad \qquad \qquad \times \; E_0[\, j_1] \cdots E_0[\, j_d]  \; \;  E\left (\, {\cal L}\, ; 
\psi [ \, j_1] \, , \cdots , \psi [\, j_d ] \right ) \qquad .
\label{iiq}
\ena

\noindent The link of Figure~2 can be understood as the connected sum of $\, ( d-1 ) \, $ Hopf
links $\, {\cal H} \, $, i.e. $ \, {\cal L} = {\cal H} \# {\cal H} \cdots \# {\cal H} \, $. 
Therefore, by using equation (\ref{somcon}), expression (\ref{iiq}) can be written as 

\beeq
I(L_{p/r}) \, = \, e^{i \theta_k \sigma ({\cal L})} \, ( \, a_k \, )^d \, \sum_{j_1, \cdots
, j_d \in \T} \, q^{\left ( \, \sum_{i=1}^d z_i \, Q(j_i) \,  \right ) } \; H_{1 j_d}\,  H_{j_d
j_{d-1}}\,  \cdots \,  H_{j_2 j_1} \,  H_{j_1 1} \quad . \label{eq:def1}
\ene 

\noindent In terms of the generators (\ref{eq:modu}) of the modular group, one finds 
\beeq
I (L_{p/r}) \; = \; e^{i \theta_k \sigma ({\cal L})} \; ( a_k )^{-1}\;  \left[ \, F(p/r)
\right]_{11}
\qquad , \label{eq:lei}
\ene
where $ \, \left[ \, F(p/r) \right]_{11} \, $ is the element corresponding to the first row and
the first column of the following matrix 
\beeq
F(p/r) \; = \; X Y^{z_d} XY^{z_{d-1}}X \cdots XY^{z_1}X \qquad . \label{matrix}
\ene
 The invariant $\, I (L_{p/r})  \, $ given in equation (\ref{eq:lei})  is in agreement with
the expressions obtained in \cite{fregomp,jef} apart from an overall normalization factor. 

\section{\bf The SU(2) case}
In this section, we shall compute $ \, I(L_{p/r}) \, $  for the gauge group $ \, G=SU(2)
\, $. Then, we will show that in this case our conjecture is true; i.e. when $ \, I(L_{p/r})
\not= 0 \, $, the absolute value $ \, | \, I(L_{p/r}) \, | \, $ only depends on $ \, p \, $. 

For $\, k \geq 2\, $, the standard basis of $\, {\cal T}_k\, $ is $\, \{ \, \psi [\, j\, ] \,
\} \, $; the index $ \, j \, $ represents the dimension of the irreducible representation
described by  $ \, \psi [\, j\, ] \,  $ and  $  \,  1 \leq j \leq (k-1) \,$. 
The matrix elements of $\, X\, $ and $\, Y\, $ are  
\bea
&&\left(X \right)_{mn} \; = \; \frac{i}{\sqrt{2k}} \left[ \exp \left(-\, \frac{i \pi mn}{k} \right)
\, - \, \exp \left(\frac{i \pi mn}{k} \right) \right] \quad ; \nb \\
&&\left(Y \right)_{mn} \; = \; \xi \, \exp \left(- \, \frac{i \pi m^2}{2k} \right) \delta_{mn} \quad
; \ena
with 
\beeq
\xi \; = \; \exp(\, i \pi/2k\, ) \qquad . 
\ene
When $ \, k=1 \, $,  one has 
\beeq
X \; = \;  \pmatrix {1 &1 \cr 1 & -1 \cr} \quad , \quad Y \; = \; \pmatrix{ 1& 0 \cr 0 & i \cr}
\quad .
\ene
The algebra $\, {\cal T}_1\, $ is isomorphic with $\, {\cal T}_3\, $ and it is easy to verify
that 
\beeq  I_{k=1}(L_{p/r})  \; =\;  \left [ \, I_{k=3}(L_{p/r}) \, \right  ]^* \qquad . 
\ene
Therefore, we only need to consider the case $\, k \geq 2\, $.  

In order to compute $ \, I(L_{p/r}) \, $, we shall derive a recursive relation for the
matrix (\ref{matrix}); the argument that we shall use has been produced by Jeffrey 
\cite{jef} in a slightly different context. In fact, our final result for $ \, I(L_{p/r}) \, $ is
essentially in agreement with the formulae obtained by Jeffrey.  Since in our approach the
invariance under Kirby moves is satisfied, our derivation of $ \, I(L_{p/r}) \, $ proves that the
appropriate expressions given in  \cite{fregomp,jef} really correspond to the values of a
topological invariant of 3-manifolds.  

Let us introduce a few definitions; with the ordered set of integers $ \{z_1, \, z_2, \cdots,
\, z_d \} $ one can define the following partial continued fraction decompositions 
\beeq
\frac{\alpha_t}{\gamma_t} \; = \;  z_t \, - \, \frac{1}{\; z_{t-1}\; -\;  \frac{1}{\; \ddots \; -
\;  \, \frac{1}{z_1}}} \qquad , 
\ene
where $ \, 1 \leq t \leq d \, $. The integers $ \, \alpha_t \, $ and $ \, \gamma_t \, $ satisfy
the recursive relations 
\bea
&&\alpha_{m+1} \; = \; z_{m+1} \, \alpha_m \, - \, \gamma_m, \quad , \quad \alpha_1 \, = \, z_1
\quad , \quad \alpha_0 \; = \; 1 \quad ;
\\ &&\gamma_{m+1} \; = \; \alpha_m  \qquad , \qquad \gamma_1 \, = \, 1  \qquad ,  
\ena
and, clearly, $\, \alpha_d / \gamma_d = p / r \, $. Finally, let $ \, F_t \, $ be the matrix 
\beeq
F_t \; = \; X Y^{z_t} XY^{z_{t-1}}X \cdots XY^{z_1}X \qquad ; \label{parma}
\ene
by definition, one has $ \, F_d = F(p/r) \, $. 

\medskip 

\no
{\bf Lemma 1}

\no
{\em The matrix element $\, (F_t)_{mn}\, $ is given by 
\bea
&&\left(F_t \right)_{mn} \; = \; B_t \, \sum_{s(m,k,|\alpha_t|) } \, \left[ \, e^{\frac{i
\pi\gamma_t }{2k \alpha_t} \left(s+\frac{n}{\gamma_t}
\right)^2} \, - \, e^{\frac{i \pi \gamma_t}{2k\alpha_t} \left(s- \frac{n}{\gamma_t}\right)^2} \,
\right]\quad ;
\label{eq:mat} \\ 
&&B_t \; = \; \frac{(-i)^{t+1}}{\sqrt{2k| \alpha_t | }} \, \xi^{z_1 + z_2 + \cdots + z_t} \, \exp
\left\{ - \frac{i \pi}{4} \left[\mbox{sign}(\alpha_0 \alpha_1) \, + \cdots +
\mbox{sign}(\alpha_{t-1} \alpha_t) \right] \right \} \nb \\
&&\qquad \null \qquad  \exp \left\{ \frac{i \pi n^2 }{2k} \left[\frac{1}{\alpha_0 \alpha_1} \, +
\cdots +
\frac{1}{\alpha_{t-2} \alpha_{t-1}} \right] \right\}\quad ; \label{bi}
\ena
where $\, s(m,k,|\alpha_t|)\, $ stands for the sum over a complete residue system modulo
$\, (2k|\alpha_t| ) \, $ with the additional constraint $\, s \equiv m \; (mod \, 2k)\, $}.

\medskip 

\no {\bf Proof}

\no The proof is based on induction. First of all we need to verify the validity of
equations (\ref{eq:mat}) and (\ref{bi}) when $\, t=1\, $. In this case, from the definition
(\ref{parma}) one gets
\beeq
\left( F_1 \right)_{mn} \; = \; - \frac{1}{2k}\,  \frac{1}{2} \, \xi^{z_1} \, \sum_{s =
1}^{2k}e^{-i  \pi s^2   z_1/(2 k)}\,  \left[\, e^{-i \pi s(m+n)/k} \, - \, e^{-i \pi s(m-n)/k} \,
+  \, \mbox{c. c.} \, \right]  \quad .
\label{eq:sum1} 
\ene
Since the sum (\ref{eq:sum1}) covers twice a complete residue system modulo $\, k \, $, i.e.  $ \,
1 \leq s \le  2k\, $,  a multiplicative factor $\, 1/2\, $ has been introduced in
(\ref{eq:sum1}). The change of variables  $\, s \rightarrow - s \, $ shows that the last two
terms in (\ref{eq:sum1}) are equal to the first two terms. Therefore, equation (\ref{eq:sum1})
can be written as 
\beeq
\left( F_1 \right)_{mn} \; = \; - \frac{1}{2k} \,  \xi^{z_1} \, \sum_{s = 1}^{2k}\,  e^{-i 
\pi s^2   z_1/(2 k)}\,  \left [ \, e^{-i \pi s(m+n)/k} \, - \, e^{-i \pi s(m-n)/k} \,  \right ] 
\label{eq:sum2}\qquad .
\ene   
At this point, one can use the reciprocity formula \cite{rec} reported in the appendix and one
gets 
\bea
&&\left( F_1 \right)_{mn} \; = \; \frac{-1}{\sqrt{2k| z_1 | }} \, \xi^{z_1} \, \exp
\left\{ - \frac{i \pi}{4} \, sign (\alpha_0 \alpha_1) \right \} \, \times \nb \\  
&&\qquad \qquad \null \qquad \times \, \sum_{v=0}^{|z_1|-1} \, \left [ \, e^{\frac{i \pi}{2k z_1}
\left ( 2 k v + m +n \right)^2} \, - \, e^{\frac{i \pi}{2k z_1} \left ( 2k v + m -n \right)^2} \,
\right] \quad . \label{inter}
\ena
By introducing the new variable $ \, s = 2k v + m \, $, one finds that in equation (\ref{inter})
the variable $\, s \, $ covers a complete residue system modulo $\, ( 2k|z_1| ) \, $ with the
constraint that $\, s \equiv m \; (mod \, 2k)\, $. Therefore, equation (\ref{inter}) can be
written in the form 
\beeq
\left( F_1 \right)_{mn} \; = \; B_1 \, \sum_{s(m,k,|z_1|)}\, 
\left [\, e^{\frac{i \pi}{2k z_1} \left(s +n \right)^2} \, - \, e^{\frac{i \pi}{2k z_1}  \left(s
-n \right)^2} \, \right] \qquad .
\ene
This confirms the validity of equation (\ref{eq:mat}) when $ \, t = 1 \, $. In order to complete
the proof, suppose now that (\ref{eq:mat}) is true for a given $\, t\, $; we shall show that 
(\ref{eq:mat}) is true also in the case $\, t+1\, $. Indeed, one has    
\beeq
\left(F_{t+1} \right)_{mn} \; = \; \sum_{v = 1}^{k} \, \left( \, X \,  Y^{z_{t+1}} \,
\right)_{mv} \, \left( \, F_t \, \right)_{vn} \qquad .
\ene
From equation (\ref{eq:mat}) one gets  
\bea
&&\left(F_{t+1} \right)_{mn} \; = \; - B_t \, \frac{i \xi^{z_{t+1}}}{\sqrt{2k}}\,  \frac{1}{2}\, 
\sum_{v =1}^{2k}\, \sum_{s(v,k,|\alpha_t|)} \, e^{- \frac{i\pi}{2k} v^2 z_{t+1}} \nb \\
&&\qquad \qquad \left[ \, e^{\frac{i \pi \gamma_t}{2 k \alpha_t} \left( s- \frac{n}{\gamma_t}
\right)^2 } \, e^{- i\pi m v/k} \, - \, e^{\frac{i \pi \gamma_t}{2 k \alpha_t} \left( s+ \frac{n}
{\gamma_t}\right)^2 }\,  e^{- i\pi mv/k} \right. \nb \\ 
&&\qquad \qquad \; \; \left. - \, e^{\frac{i \pi \gamma_t}{2 k \alpha_t} \left( s-
\frac{n}{\gamma_t} \right)^2 }\,  e^{ i \pi mv/k} \, + \, e^{\frac{i \pi \gamma_t}{2 k \alpha_t}
\left(s +\frac{n}{\gamma_t}
\right)^2 } \, e^{ i \pi mv/k} \right ] \quad .
\ena
Again, the last two terms can be omitted provided one introduces a multiplicative factor $ \, 2
\, $. Moreover, because of the constraint $\, v = s \, (mod \, 2k)$, one can set $\, v =
s\, $, thus
\bea
&&\left(F_{t+1} \right)_{mn} \; = \; - B_t \,  \frac{i \, \xi^{z_{t+1}}}{\sqrt{2k}} \, e^{\frac{i
\pi n^2}{2k \alpha_t \gamma_t}}\, \sum_{s = 0}^{2k|\alpha_t|-1} \nb \\
&&\qquad \left \{ \, e ^{-\frac{i \pi}{2k \alpha_t} \left[ \alpha_{t+1} s^2 + 2 \left(\gamma_{t+1}
m + n \right)s \right]} \, - \, e ^{-\frac{i \pi}{2k \alpha_t} \left[ \alpha_{t+1} s^2 + 2
\left(\gamma_{t+1} m - n
\right) s \right]} \, \right \} \quad .
\ena
By using the reciprocity formula, one obtains the final expression for $\, (F_{t+1})_{mn}\, $
\bea
&&(F_{t+1})_{mn}\; = \; - i \, B_t \,  \xi^{z_{t+1}}\,  \sqrt{\frac{|\alpha_t|}{|\alpha_{t+1}|} }
\, e^{\left(\frac{i \pi n^2}{2k \alpha_t \alpha_{t+1}} \right)} \, e^{ \frac{- i \pi}{4}{ 
sign}(\alpha_t \alpha_{t+1})} \nb \\ 
&& \qquad \; \sum_{v=1}^{|\alpha_{t+1}|} \, \left \{ \, e^{\frac{i \pi \alpha_t}{2k\alpha_{t+1}}
\left ( \, 2kv \, + \, m \, + \, \frac{n}{\alpha_t} \, \right)^2 } \, - \,   
e^{ \frac{i \pi \alpha_t}{2k\alpha_{t+1}} \left(\, 2kv \, + \, m \, - \, \frac{n}{\alpha_t}\, 
\right)^2 } \, \right \} \quad . \label{final}
\ena
In terms of the variable $ \, s = 2kv + m \, $, equation (\ref{final}) can be rewritten in the
form (\ref{eq:mat}) and this concludes the proof. {\hfill $\spadesuit$}

\medskip 

From the definition (\ref{eq:lei}) and {\bf Lemma 1} it follows 

\no {\bf Theorem 1}

\no {\em Let $SU(2)$  be the gauge group, the 3-manifolds invariant $\, I_k (L_{p/r})\, $ for
$\, k \geq 2\, $ is given by}
\bea
&&I_k(L_{p/r}) \; = \;   \sum_{s \, (mod \;  p)} \left \{\, \exp{ \left[ \frac{i
\pi (r+1)^2}{2p k r} \right]}  \,  \exp \left [ \frac{i 2 \pi}{p}
\left[r ks^2 \, + \,  (r+1)s \right]  \right] \right.   \nb \\
&& \quad - \left. \, \exp{\left[ \frac{i \pi (r-1)^2}{2p k r} \right]}
 \, \exp \left[{ \frac{i 2 \pi}{p}
\left[r ks^2 \,  + \,  (r -1)s \right]} \right] \right \} \, \frac{e^{i
\theta_k \sigma ({\cal L}) }\> B_d}{ a_k}
\,  \quad . \label{invlens}
\ena

\medskip 

\no {\bf Proof}

\no According to equation (\ref{eq:lei}), the expression for the matrix element $ \, \left[ \,
F(p/r) \right]_{11} \, $ has been written by means of a sum over a complete residue system
modulo $\, p \, $. {\hfill $\spadesuit$} 

\medskip 

As shown in equation (\ref{invlens}),  the expression for $\, I_k ( L_{p/r} ) \, $ is
rather involved; nevertheless, $\, | \,  I_k ( L_{p/r} ) \, |^2 \, $ can be computed
explicitly. Let us introduce the modulo-$ p\, $ Croneker delta symbol defined by  
\beeq
\delta_p \left ( x \right) \; = \; \left \{ \begin{array}{cc} 0 \qquad & x \, \not \equiv \,  0 \; \;
( \,  mod \, p \, ) 
\\ 1 \qquad & x \equiv 0 \; \; (\,  mod \, p \, )  
\end{array} \right. \qquad ;
\ene
where $\, p \, $ and $\, x\, $ are integers. One can easily verify that, for integer $ \, n \,
$,  
\beeq 
\left \{ \begin{array}{cc} \delta_p(xn) \; = \; \delta_p(x) \qquad  &{\rm if } \quad  (n,p) \, = \, 1
\quad ; \\
\delta_{pn}(xn) \; = \; \delta_p(x)  \qquad . & \qquad \end{array}\right. 
 \label{proper} 
\ene
Finally, we shall denote by $\, \phi (n) \, $ the Euler function \cite{cin} which is equal to the
number of residue classes modulo $\, n \, $ which are coprime with $ \, n \, $. 

\medskip 

\no
{\bf Theorem 2}

\no
{\em The square of the absolute value of $\, I_k  ( L_{p/r}  ) \, $ is given in the following
list~; 

\no for $\, p =2\, $
\beeq
\left | \, I_k  ( L_{2/1} ) \, \right |^2 \; =  \; \left[ \, 1+(-1)^k \, \right] \, 
\frac{\sin^2 \left[\pi/(2k) \right]}{\sin^2 \left[\pi/k \right]} \qquad ; \label{facile} 
\ene
for $\, p  > 2\, $ one has : 

\no when $\, p \, $ and $ \, k \, $ are coprime integers, i.e. $\, (k,p) \, = \, 1\, $,  
\bea 
&& \left | \, I_k ( L_{p/r} ) \, \right |^2 \; =  \; \frac{1}{2}\, \left[ \, 1-(-1)^p \, \right]
\, \frac{ \sin^2 \left[\, \pi \left( \, k^{\phi(p)} -1 \, \right)/(kp) \, \right] }{\sin^2
(\pi/k)}  \, + \nb \\ 
&& \qquad \frac{1}{2}\, \left[\, 1+(-1)^p \, \right] \, \left[ \, 1+(-1)^{p/2} \, \right] \, 
\frac{  \sin^2 \left[ \, \pi \left( \, k^{\phi(p/2)}-1\, \right)/(kp) \right] }{\sin^2(\pi/k)}
\qquad ; 
\ena
when the greatest common divisor of $ \, p \, $ and $\, k \, $ is greater than unity, i.e. $\,
(k,p) \, = \, g >1\, $ and $\, p /g \, $ is odd
\beeq
\left | \, I_k ( L_{p/r} )\, \right |^2  \; =  \; \frac{g}{4\, \sin^2( \pi/k) } \, 
\left[\, \delta_g(\, r-1\, ) \, + \, \delta_g(\, r+1\, ) \, \right] \qquad ; \label{44}
\ene
when $\, (k,p) \, = \, g >1 \, $ and $\, p /g \, $ is even
\bea
&& \left | \, I_k ( L_{p/r} ) \, \right |^2 \; =  \; \frac{g}{4\, \sin^2( \pi/k) } \, 
\left \{ \, \delta_g(r+1) \, \left [ \, 1 + (-1)^{kp/2g^2} \, (-1)^{(r+1)/g} \, \right ] \, +
\right. \nb \\  
&& \left. \qquad \delta_g(r-1) \, \left [ \, 1 + (-1)^{kp/2g^2} \, (-1)^{(r-1)/g} \, \right ] \,
\right \} \qquad . \label{45}
\ena }

\medskip

\no
{\bf Proof}

\no From Theorem~1 it follows that the square of the absolute value of the lens space invariant is 
\beeq
\left |\, I_k (L_{p/r} )\,  \right |^2 \; =  \; a(k)^{-2} \, \left( 2kp \right)^{-1} \,
{\cal S}(k,p,r) \qquad , 
\ene
with
\bea
&&{\cal S}(k,p,r) \; = \;  \sum_{s,t \, \left(mod \, p \right)} \left \{ \, \exp \left\{\frac{i 2
\pi}{p} \left[ kr \left(s^2-t^2 \right) \, + \, \left(r+1 \right) \left(s-t \right) \right] \right \}
\right. \nb \\ 
&&\qquad - \, \exp\left(\frac{i 2 \pi}{kp} \right) \, \exp \left\{\frac{i 2 \pi}{p}
\left[ kr \left(s^2-t^2 \right) \, + \, r \left( s-t \right) \, +\, s\,  +\, t\, \right]
\right \} \nb \\
&& \qquad - \exp\left(- \frac{i 2 \pi}{kp} \right) \, \exp \left\{\frac{i 2 \pi}{p}
\left[ kr \left(s^2-t^2 \right) \, + \, r \left( s-t \right) \, - \, s \, - \, t \,  \right]
\right
\} \nb \\
&& \qquad \qquad + \left. \, \exp \left \{ \frac{i 2 \pi}{p} 
\left[ kr \left(s^2-t^2 \right) \, + \, \left( r-1 \right) \left( s-t \right) \right] \right \} 
\right \} \qquad . \label{123}
\ena
The indices  $\, s\, $ and $ \, t \, $ run over a complete residue system modulo $\, p\, $. When $\,
p \, = \, 2\, $,  each sum contains only two terms and the evaluation of (\ref{123}) is
straightforward; the corresponding result is shown in equation (\ref{facile}). Let us now consider
the case in which $ \, p > 2 \, $. By means of the change of variables $\, s \rightarrow s+t\, $, 
the sum in $\, t\, $ becomes a geometric sum and one obtains 
\bea 
&&{\cal S}(k,p,r) \; = \; p \,  \sum_{s \, \left(mod \, p \right)} \left \{ \exp \left\{\frac{i 
2 \pi}{p} \left[ kr s^2  \, + \, \left(r+1 \right)s \right ] \right \}  \, \delta_p \left(2krs
\right) \right. \nb \\
&&\qquad - \, \exp \left(\frac{i 2 \pi}{kp} \right) \, \exp \left\{\frac{i 2 \pi}{p}
\left[ krs^2 \, + \, \left(r+1 \right)s  \right] \right \} \, \delta_p (2krs \, + \, 2)
\nb
\\
&&\qquad - \, \exp\left(\frac{- i 2 \pi}{kp} \right) \, \exp \left\{\frac{i 2 \pi}{p}
\left[ krs^2 \, + \, \left(r-1 \right)s  \right] \right \}  \, \delta_p (2krs \, - \, 2) \nb
\\
&&\qquad + \, \left.   \exp \left\{\frac{i 2 \pi}{p}
\left[ kr s^2  \, + \, \left(r-1 \right)s \right ] \right \} \, \delta_p \left(2krs \right)
\right \} \qquad . \label{eq:sum10}
\ena
By using properties (\ref{proper}), one can determine the values of $\, s\, $
which give contribution to (\ref{eq:sum10}). Let us start with $\, (k,p) \, =  \, 1\, $.  Clearly,
in this case   one has  
\beeq
\delta_p(2rks) \, \neq \, 0 \Rightarrow \left \{ \begin{array}{cc} s \, = \, p & \qquad p \; \mbox{
odd} \\ s \, = \, p, \, p/2 & \qquad p \; \mbox{ even} \end{array}  \right. \quad . 
\ene
When $\, (k,p) \, =  \, 1\, $ and $\, p\, $ is odd, one gets
\beeq
\delta_p (2krs \; \mp \; 2) \, = \, \delta_p (krs \; \mp \; 1) \qquad .
\ene
The delta gives a non-vanishing contribution if and only if the following congruence is satisfied
\beeq
rks \, = \, \pm 1 \; \; (mod \, p) \quad . \label{eq:con}
\ene
The unique solution \cite{cin} to (\ref{eq:con}) is given
 by 
\beeq
s \; = \; \pm {(rk)}^{\phi(p)-1} \qquad . 
\ene
When $\, (k,p) \, =  \, 1\, $ and $ \, p \, $ is even, one finds two solutions 
\beeq 
s_1 \; = \; \pm {(rk)}^{\phi(p/2)-1} \quad , \quad  s_2 \; = \; \pm  {(rk)}^{\phi(p/2)-1} \, +
\, p/2   \qquad . 
\ene
Let us now examine the case $\, (p,k)  =   g > 1\, $. We introduce the integer $\, \beta \, $ defined
by $\, p  = g \beta \, $. For $\, \beta \, $ odd, one has
\beeq
\delta_p(2krs) \;  = \; \delta_\beta (s) \qquad .
\ene
Within the  residues of a complete system modulo $\, p\, $, the values of $\, s\, $ giving
non-vanishing contribution are of the form $\, s = \alpha \beta \, $ with $ \, 1 \leq \alpha
\leq g\, $. When $\, \beta \, $ is even, one gets 
\beeq
\delta_p(2krs) \;  = \; \delta_\beta (2s) \; = \; \delta_{\beta /2} (s) \qquad .
\ene
The solutions of the associated congruence are 
\beeq
s \; = \; \alpha \, \frac{\beta}{2} \qquad  1 \leq \alpha \leq 2g \qquad .   
\ene   
When $\, (k,p) \, = \, g >1\, $ and $\, p\, $ is odd, $\, \delta_p \left[2r(ks \pm 1) \right]\, $
does not contribute because $\, rks  =  \pm 1 \; (mod \, p)\, $ has no solutions. On the other hand,
if $\, p\, $ is even we have
\beeq
\delta_p \left [ \, \left( 2rks \; \pm \; 2 \right )\,  \right] \; = \; \delta_{p/2}(rks \pm 1)
\qquad . \label{134}
\ene
The delta function (\ref{134}) is non-vanishing when $\, (p/2,k) \, = \, 1\, $ and, in this case, 
the two solutions are $\, s_1  =   \pm {(rk)}^{\phi(p/2)-1}\, $ and $ \, s_2 = s_1 +
p/2 \, $.  This exhausts the analysis of the modulo $\, p\, $ Croneker deltas when $\, p >2\, $. 

\no
At this stage, Theorem 2 simply follows from the substitution of the values of $\, s \,$ for which
the various Croneker deltas modulo $ \, p \, $ are non vanishing. In the case $ \, (k,p) =1 \, $ and
$\, p \, $ odd, the algebraic manipulations are straightforward. When $ \, (k,p) =1 \, $ and
$\, p \, $ even, the evaluation of (\ref{eq:sum10}) needs some care. In this case, one has to deal
with factors of the form 
\beeq
\exp \left[\frac{i \pi}{b} \left (a^{\phi(b)} -1 \right) \right ] \qquad ; \label{eli}
\ene
with $\, b > 2 \, $ even and $\, (a, b)   = 1 \, $. In appendix B, it is shown that terms of the
type  (\ref{eli}) are trivial because actually
\beeq
a^{\phi(b)} \; \equiv \; 1 \quad (\, mod \; 2b) \qquad .
\ene
Finally, the derivation of equations (\ref{44}) and (\ref{45}) is straightforward.   {\hfill
$\spadesuit$}

\medskip

\no
Let us now consider the dependence of $\, | \, I (L_{p/r})\, |^2 \, $ on $ \, r \, $. As shown in
equations (\ref{44}) and (\ref{45}), $\, | \, I (L_{p/r})\, |^2 \, $ depends on $\, r \, $.
However, this dependence is rather peculiar: when $ \, I (L_{p/r}) \not= 0 \,  $, $\, | \, I
(L_{p/r})\, |^2 \, $ does not depend on $\, r \, $. Indeed, when expression
(\ref{44}) is different from zero, its values are given by 
\beeq 
0 \; \not= \; \mbox{(\ref{44})} \; = \; \left \{ \begin{array} {cc} \sin^{-2}(\pi /k) & \quad \mbox{
for}
\; g =2
\; ;
\\  (g/4) \, \sin^{-2}(\pi/k)  & \quad \mbox{ for} \; g >2 \; . 
\end{array} \right.
\ene
Similarly, when expression (\ref{45}) is different from zero, its value is given by
\beeq
0 \; \not= \; \mbox{(\ref{45})} \; = \; \frac{g}{2 \, \sin^2(\pi/k)} \qquad .
\ene 
To sum up, when $ \, I (L_{p/r}) \not= 0 \,  $, $\, | \, I (L_{p/r})\, |^2 \, $ only depends on 
$\, p\, $ and, therefore, it is a function of the fundamental group $\, \pi_1(L_{p/r})\, =
\, Z_p \, $. Thus, Theorem~2 proves the validity of our conjecture for the lens
spaces when the gauge group is $\, SU(2)\, $.

\section{\bf The SU(3) case}
In this section we shall present numerical computations confirming the validity of our conjecture for
lens spaces when the gauge group is $ \, SU(3) \, $.  As in the $\, SU(2)\, $ case, the $\, SU(3)\, $
Chern-Simons field theory can be  solved explicitly in any closed, connected and orientable
three-manifold \cite{gp1}.  The general surgery rules for $ \, SU(3) \, $ and for any integer
 $ \, k \, $ have been derived in \cite{gp1}. In particular, it turns out that 
\beeq  
I_{k=1}(L_{p/r})  \; =\;  \left [ \, I_{k=2}(L_{p/r}) \, \right  ]^* \; = \; I_{k=4}(L_{p/r}) 
\qquad . 
\ene
Therefore, we only need to consider the case $\, k \geq 3\, $.  For $ \, k \geq 3 \,
$, the matrices  which give a projective representation  of the modular group have the following form
\bea
&&X_{(m,n) \, (a,b)} \; = \;  \frac{i}{k \sqrt{3}} q^{-2} q^{-[ (m+n) (a+b+3) + (m+3) b + (n+3)
a] /3}  \nb \\
&&\qquad  \left[ \, 1 + q^{(n+1)(a+b+2)+(m+1)(b+1)} + q^{(m+1)(a+b+2) + (n+1)(a+1)}  \right. \nb \\
&&\qquad \left. - \, q^{(m+1)(b+1)} \, - \, q^{(n+1)(a+1)} \, - \, q^{(m+n+2)(a+b+2)} \right] \quad ;
\ena
\beeq
Y_{(a,b) \, (m,n)} \; = \; q^{[m^2 + n^2 + m n + 3 (m+n)]/3} \delta_{am} \, \delta_{bn}
\qquad ;
\ene
\beeq 
C_{(a,b) \, (m,n)} \; = \; \delta_{an} \delta_{bm} \qquad ;
\ene
where each irreducible representation of $\, SU(3)\, $ has been denoted by  a
couple of nonnegative integers $ \, (m,n)\, $  (Dynkin labels).  

\no
By using equation (\ref{eq:def1}), we have computed $\, I_k(L_{p/r})\, $ numerically for some
examples of lens spaces. In particular, we have worked out the value of the invariant for the lens
spaces $\, L_{p/r}\, $, with $\, p \leq 20\, $ and $\, 3 \leq k \leq 50 \, $.  In all these cases,
the results are in agreement with our conjecture. 

Our calculations have been performed on a Pentium based PC running Linux. 
For instance, the results of the computations for the cases $\, L_{8/1}\, ,
\, L_{8/3}\, , \, L_{15/1}\, , \, L_{15/2}\, , \, L_{15/4} \, $ with $\, 3 \leq k \leq 50 \, $ are
shown in Tables 1, 2, 3, 4, 5.   The spaces $\, L_{8/1}\, $ and $\, L_{8/3}\, $ are not homotopically
equivalent; as shown in Tables~1 and 2, the phase of the invariant distinguishes these
two manifolds.  The case in which $ \, p=15 \, $ is more interesting because there are two different
spaces belonging to the same homotopy class;  $\, L_{15/1}\, $ and $\, L_{15/4}\, $ are
homotopically equivalent and $\, L_{15/2}\, $ represents the other homotopy class. The phase
of the invariant distinguishes the manifolds of the same homotopy class. 

\section{\bf Conclusions}

In this article, we have presented some arguments and numerical results supporting the conjecture
that, for nonvanishing  $\, I(M) \, $, the absolute value $\, | \,  I(M) \, | \, $ only depends on
the fundamental group $\pi_1 (M) \, $. Since the Turaev-Viro invariant \cite{vitur} coincides
\cite{tur} with  $\, | \,  I(M) \, |^2 \, $,  our conjecture gives some hints on the   
topological interpretation of the Turaev-Viro invariant.  For the gauge group $ \, SU(2) \, $,   
$\, | \,  I(M) \, |^2 \, $ can be understood as the improved partition function of the Euclidean
version of (2+1) gravity with positive cosmological constant \cite{arwil,tom}. In this case, our
conjecture  suggests that, for instance,  the semiclassical limit is uniquely determined by the
fundamental group of the universe.  

Finally, one may ask for which values of $ \, k \, $ the equality  $\, I_k(M) =0\, $ is  satisfied
and what the meaning of this fact is. The complete solution to this problem is not known. From the
field theory point of view,  gauge invariance of the factor $\, \exp \left ( i S_{_{CS}} \right ) \,
$ (where $\, S_{_{SC}} \, $ is the  Chern-Simons action) in the functional measure gives nontrivial
constraints on the admissible values of $ \, k \, $ in a given manifold $ \, M \, $. In certain
cases \cite{gua3} one finds that, in correspondence with the ``forbidden" values of  $\, k \, $, the
invariant $\, I_k(M)\, $ vanishes. So, it is natural to expect that  $\, I_k(M) =0\, $ is related to
a breaking of gauge invariance for large gauge transformations. From the mathematical point of view,
$\, I_k(M) =0\, $ signals the absence of the natural extension of $\, E( {\cal L})  \, $ to an
invariant $\, E_{_{M}}( {\cal L})  \, $ of links in the manifold $ \, M \, $. More precisely, when 
$\, I_k(M) \not= 0\, $ for fixed integer $ \, k \, $, one can define \cite{gua2} an invariant  $\,
E_{_{M}}( {\cal L})  \, $ of oriented, framed and coloured links $ \, \{ \, {\cal L} \subset M \,
\}\, $ with the following property: if the link $ \, {\cal L} \, $ belongs to a three-ball embedded
in $ \, M \, $, then one has   $\, E_{_{M}}( {\cal L})  = E( {\cal L}) \, $. The values of
the invariant $\, E_{_{M}}( {\cal L})  \, $ correspond to the vacuum expectation values of the 
Wilson line operators associated with links in the manifold $ \, M \, $. When $\, I_k(M) =0\, $,
the invariant $\, E_{_{M}}( {\cal L})  \, $ cannot be constructed; consequently, for these 
particular values of $ \, k \, $,   the quantum Chern-Simons field theory is  not well
defined in $ \, M \, $. 

\vskip 1.5 truecm 

\no {\bf Acknowledgments.} ~We wish to thank Turaev for useful discussions. 

\vskip  2 truecm

\no {\bf  Appendix A}

\no The generalized Gauss sums have a very useful property which can be
expressed by means of the so-called reciprocity formula  \cite{rec} 
\beeq
\sum_{n=0}^{|c|-1}\, e^{i \frac {\pi}{c}(an^2+bn)}\; = \; \sqrt{\left| \frac{c}{a}
\right|}\; e^{i \frac{\pi}{4ac}(|ac|-b^2)} \; \sum_{n=0}^{|a|-1}e^{-i \frac{\pi}{a}
(cn^2+bn)} \qquad ,  
\ene
where the integers $\, a, \, b, \, c\, $ satisfy the relations 
\beeq
 ac \, \neq \; 0 \quad \; \qquad , \qquad   ac \; + \; b \mbox{ ~is even} \qquad . 
\ene

\vskip 1truecm

\no{\bf  Appendix B}

\no {\bf Lemma~2} ~{\em Let $\, a, \, b \, $ two integers, with $\, (a,b) = 1 \, $ and $\, b >2 \, 
$ even;  one  has}
\beeq
a^{\phi(b)} \, \equiv \, 1 \quad (\, mod \; 2b) \qquad . \label{999} 
\ene
{\bf Proof}

\no The proof consists of two parts: firstly, it is shown by induction that Lemma~2 holds when $\, b
= 2^m\, $ with $\, m >1 \, $ integer. Secondly, equation (\ref{999})  is proved when $\, b = 2^m \,
c \, $ with   $\, m \geq 1 \, $  and $\, c \, $ odd integer. 

Since $\, b \, $ is even, $\, a \, $ is clearly odd and can be written in the form $\, a = (2f +1)
\, $. When $ \, b \, $ is of the type $\, b = 2^m\, $, the condition $ \, b >2 \, $ implies that $
\, m \geq 2 \, $. Let us now consider the case   $\, m = 2 \, $; one has $ \, \phi (b) = \phi (\,
2^2\, ) = 2\, $, therefore 
\beeq
a^{\phi(b)} \; = \; (2f \; + \; 1)^2 \; = \; 1 \, + \, 4 f ( \, f+1 \, )  \; \equiv \; 1 \quad (\,
mod \; 2^3 \, ) \qquad .
\ene
Thus, {\bf Lemma~2} is satisfied when $\, b= 2^2 \, $. Suppose now that equation (\ref{999})
holds when $\, b = 2^n \,$ for a certain $ \, n \, $.  We need to prove that  (\ref{999}) is true
also for  $\, b= 2^{(n +1 )}  \,$. Indeed, $\, \phi(2^{n+1}\, ) = 2^n\, $ and one gets
\beeq
(2f \; + \; 1)^{\phi(2^{n+1})} \; = \; \left[\, (2f \;+ \; 1 )^{\phi ( 2^n \, ) } \, \right]^2
 \qquad .
\ene
By using the induction hypothesis
\beeq 
(2f \;+ \; 1 )^{\phi ( 2^n \, ) } \; = \; 1 \, + \, N \, 2^{n+1} \qquad , 
\ene
one finds 
\beeq
\left[\, (2f \;+ \; 1 )^{\phi ( 2^n \, ) } \, \right]^2 \; = \; 1 \, + \, 2^{n+2} \, N \,
\left(\, 1\, + \, 2^n \, N \, \right) \; \equiv \; 1 \quad (\, mod \; 2^{n+2} \,) \; .
\ene
Therefore, equation  (\ref{999}) is also satisfied when $\, b= 2^{(n +1 )}  \,$. To sum up, for $
\, m > 1 \, $ and $ \, a \,  $ odd, one has 
\beeq
a^{\phi(2^m)}  \; \equiv \; 1 \quad (\, mod \; 2^{m+1} \, ) \qquad . \label{first}
\ene
Let us now consider the general case in which $\, b = 2^m \, c \, $ with $\, c \, $ odd integer. 
From Euler Theorem \cite{cin} it follows that 
\beeq
a^{\phi(b)} \; \equiv \; 1 \quad (\, mod \; b) \, \quad  \Rightarrow \,  \quad a^{\phi(b)}  \;  
\equiv \; 1 \quad (\, mod \; c) \quad .
\label{part1} 
\ene
On the other hand, $ \, \phi( 2^m  c \, ) =  \phi( 2^m \, ) \phi (  c  ) \, $ and, for $ \, m > 1
\, $,  equation  (\ref{first}) implies
\beeq
a^{\phi(b)} \; = \; a^{ \phi(c)\, \phi(2^m\, )} \; \equiv \; 1 \quad (mod \; 2^{m+1}) \qquad . 
\label{part2} 
\ene
Since $\, (2^{m+1}, c)  =1 \, $, from equations (\ref{part1}) and (\ref{part2}) one gets
\beeq
a^{\phi(b)} \; \equiv \; 1 \quad (mod \; 2^{m+1} \, c ) \; \equiv \; 1 \quad (mod \; 2b )\qquad .
\ene
Finally, we need to consider the case $ \, b= 2 \, c \, $. Since $ \, \phi (c) \, $
is even, one gets 
\beeq
a^{\phi (2c)} \; = \; \left [ \, 1 \, + \, 4f(f+1) \, \right]^{\phi(c)/2} \; \equiv \; 1 
\quad (\, mod \; 2^{2} \, ) \qquad . \label{fine}
\ene
Equations (\ref{first}) and (\ref{fine}) imply 
\beeq
a^{\phi (2c)} \; \equiv \; 1 \quad (\, mod \; 2^{2}c\, ) \qquad . 
\ene
This concludes the proof. {\hfill $\spadesuit$}

\textwidth =  6.0 in
\textheight = 11 in
\oddsidemargin = 0.1 cm
\topmargin = -2.2 cm
\pagestyle{empty}

\centerline{
\begin{tabular}{|c|c|c|}
\hline
$\bullet$ & $L_{8/1}$ & $\bullet$ \\ \hline
$k$ & $I_k$ & $|I_k|$ \\ \hline
3 & 1.000000000 -  i 0.000000175 & 1.000000000 \\ \hline
4 & -1.000000000 +  i 0.000000012 & 1.000000000 \\ \hline
5 & -0.499999839 +  i 1.538841821 & 1.618033989 \\ \hline
6 & -2.000000000 +  i 0.000000175 & 2.000000000 \\ \hline
7 & -1.000000000 -  i 0.000000095 & 1.000000000 \\ \hline
8 & -6.828427084 +  i 6.828427165 & 9.656854249 \\ \hline
9 & -0.499999950 +  i 0.866025433 & 1.000000000 \\ \hline
10 & -4.236067816 +  i 3.077683759 & 5.236067977 \\ \hline
11 & -2.073846587 -  i 14.423920506 & 14.572244935 \\ \hline
12 & -7.464102180 -  i 12.928202904 & 14.928203230 \\ \hline
13 & -12.373524802 -  i 17.926145664 & 21.781891892 \\ \hline
14 & 18.195669358 +  i 0.000000868 & 18.195669358 \\ \hline
15 & 5.657005398 +  i 4.110054701 & 6.992443043 \\ \hline
16 & 63.431390926 +  i 26.274142180 & 68.657642707 \\ \hline
17 & 6.721172941 +  i 2.603796085 & 7.207906752 \\ \hline
18 & 15.581719525 +  i 26.988328071 & 31.163437478 \\ \hline
19 & -69.185356387 -  i 11.544994773 & 70.142001987 \\ \hline
20 & -66.118464248 +  i 0.000001734 & 66.118464248 \\ \hline
21 & -90.016155013 +  i 0.000000715 & 90.016155013 \\ \hline
22 & 43.367008373 -  i 50.048195295 & 66.223253224 \\ \hline
23 & 22.973052202 -  i 3.157573698 & 23.189036183 \\ \hline
24 & 219.054514271 -  i 58.695485329 & 226.781922164 \\ \hline
25 & 23.000602198 -  i 5.905551129 & 23.746646829 \\ \hline
26 & 84.434763344 +  i 44.314782640 & 95.357376335 \\ \hline
27 & -185.409752744 +  i 67.483626877 & 197.308936212 \\ \hline
28 & -161.491033989 +  i 77.769978391 & 179.241523084 \\ \hline
29 & -214.342425435 +  i 99.165358161 & 236.170369862 \\ \hline
30 & 50.544948881 -  i 155.561366312 & 163.566899299 \\ \hline
31 & 47.834849646 -  i 26.550506056 & 54.709251617 \\ \hline
32 & 443.615385766 -  i 296.414325177 & 533.531688523 \\ \hline
33 & 46.894822945 -  i 30.137471278 & 55.743982582 \\ \hline
34 & 211.662329441 +  i 39.566544008 & 215.328709440 \\ \hline
35 & -344.610722364 +  i 250.374335362 & 425.962272715 \\ \hline
36 & -290.376668930 +  i 243.654963011 & 379.059824907 \\ \hline
37 & -381.345093602 +  i 307.914657384 & 490.138262785 \\ \hline
38 & 27.064369107 -  i 326.618342223 & 327.737732877 \\ \hline
39 & 79.852315000 -  i 70.742976321 & 106.681586554 \\ \hline
40 & 734.287201006 -  i 734.287220264 & 1038.438931957 \\ \hline
41 & 78.053570702 -  i 75.119028091 & 108.329258655 \\ \hline
42 & 408.590935390 -  i 0.000001624 & 408.590935390 \\ \hline
43 & -545.541472393 +  i 565.843251992 & 785.998781122 \\ \hline
44 & -452.077098514 +  i 521.724781996 & 690.340822456 \\ \hline
45 & -590.050989168 +  i 655.318028318 & 881.817377951 \\ \hline
46 & -39.324647251 -  i 574.905929557 & 576.249299975 \\ \hline
47 & 118.947578576 -  i 140.691854636 & 184.235513433 \\ \hline
48 & 1090.316030520 -  i 1420.927547540 & 1791.039960963 \\ \hline
49 & 116.363242798 - i  145.914887153 & 186.632147733 \\ \hline
50 & 687.196328608 -  i 86.813088226 & 692.658145364 \\ \hline
\end{tabular}
}
\vskip 0.2truecm
\centerline{\bf Table 1}

\centerline{
\begin{tabular}{|c|c|c|}
\hline
$\bullet$ & $L_{8/3}$ & $\bullet$ \\ \hline
$k$ & $I_k$ & $|I_k|$ \\ \hline
3 & 1.000000000 +  i 0.000000000 & 1.000000000 \\ \hline
4 & -1.000000000 +  i 0.000000000 & 1.000000000 \\ \hline
5 & 1.309016994 -  i 0.951056516 & 1.618033989 \\ \hline
6 & -2.000000000 +  i 0.000000000 & 2.000000000 \\ \hline
7 & -0.623489802 +  i 0.781831482 & 1.000000000 \\ \hline
8 & 0.000000000 +  i 0.000000000 & 0.000000000 \\ \hline
9 & -0.500000000 +  i 0.866025404 & 1.000000000 \\ \hline
10 & 1.618033989 +  i 4.979796570 & 5.236067977 \\ \hline
11 & 6.053529319 -  i 13.255380237 & 14.572244935 \\ \hline
12 & -7.464101615 +  i 12.928203230 & 14.928203230 \\ \hline
13 & 7.723965314 -  i 20.366422715 & 21.781891892 \\ \hline
14 & -16.393731622 +  i 7.894805057 & 18.195669358 \\ \hline
15 & -2.160783733 +  i 6.650208521 & 6.992443043 \\ \hline
16 & 0.000000000 +  i 0.000000000 & 0.000000000 \\ \hline
17 & -1.972537314 +  i 6.932749548 & 7.207906752 \\ \hline
18 & 15.581718739 +  i 26.988328525 & 31.163437478 \\ \hline
19 & 17.218843527 -  i 67.995675380 & 70.142001987 \\ \hline
20 & -20.431729095 +  i 62.882396270 & 66.118464248 \\ \hline
21 & 20.030478885 -  i 87.759262069 & 90.016155013 \\ \hline
22 & -55.710545730 +  i 35.802993757 & 66.223253224 \\ \hline
23 & -4.717948848 +  i 22.704016336 & 23.189036183 \\ \hline
24 & 0.000000000 +  i 0.000000000 & 0.000000000 \\ \hline
25 & -4.449677900 +  i 23.326028428 & 23.746646829 \\ \hline
26 & 54.169163837 +  i 78.477582217 & 95.357376335 \\ \hline
27 & 34.262337211 -  i 194.311370120 & 197.308936212 \\ \hline
28 & -39.884991120 +  i 174.747563877 & 179.241523084 \\ \hline
29 & 38.208113963 -  i 233.059184818 & 236.170369862 \\ \hline
30 & -132.328401250 +  i 96.142211171 & 163.566899299 \\ \hline
31 & -8.284500381 +  i 54.078362271 & 54.709251617 \\ \hline
32 & 0.000000000 +  i 0.000000000 & 0.000000000 \\ \hline
33 & -7.933195866 +  i 55.176589215 & 55.743982582 \\ \hline
34 & 129.764538515 +  i 171.836019661 & 215.328709440 \\ \hline
35 & 57.178306982 -  i 422.107212669 & 425.962272715 \\ \hline
36 & -65.823047822 +  i 373.301054424 & 379.059824907 \\ \hline
37 & 62.256293514 -  i 486.168356194 & 490.138262785 \\ \hline
38 & -258.631121471 +  i 201.300681961 & 327.737732877 \\ \hline
39 & -12.859044288 +  i 105.903757675 & 106.681586554 \\ \hline
40 & 0.000000000 +  i 0.000000000 & 0.000000000 \\ \hline
41 & -12.423570453 +  i 107.614511930 & 108.329258655 \\ \hline
42 & 254.752281347 +  i 319.449256739 & 408.590935390 \\ \hline
43 & 85.965636475 -  i 781.283554973 & 785.998781122 \\ \hline
44 & -98.245742501 +  i 683.314148273 & 690.340822456 \\ \hline
45 & 92.175015400 -  i 876.986690089 & 881.817377951 \\ \hline
46 & -447.003088251 +  i 363.663986140 & 576.249299975 \\ \hline
47 & -18.441181139 +  i 183.310248618 & 184.235513433 \\ \hline
48 & 0.000000000 +  i 0.000000000 & 0.000000000 \\ \hline
49 & -17.920983557 +  i 185.769741658 & 186.632147733 \\ \hline
50 & 441.516918550 +  i 533.702273720 & 692.658145364 \\ \hline
\end{tabular}
}
\vskip 0.2truecm
\centerline{\bf Table 2}

\pagebreak
\centerline{
\begin{tabular}{|c|c|c|}
\hline
$\bullet$ & $L_{15/1}$ & $\bullet$ \\ \hline
$k$ & $I_k$ & $|I_k|$ \\ \hline
3 & 1.000000000 -  i 0.000000175 & 1.000000000 \\ \hline
4 & 0.000000021 +  i 1.732050808 & 1.732050808 \\ \hline
5 & -2.665351925 +  i 1.936491953 & 3.294556414 \\ \hline
6 & 3.000000303 +  i 3.464101353 & 4.582575695 \\ \hline
7 & -0.000000165 +  i 1.732050808 & 1.732050808 \\ \hline
8 & -1.732050808 +  i 0.000000010 & 1.732050808 \\ \hline
9 & -0.907604426 -  i 11.866568847 & 11.901226911 \\ \hline
10 & -5.959909504 -  i 18.342712091 & 19.286669182 \\ \hline
11 & -17.245203220 +  i 14.943053456 & 22.818673947 \\ \hline
12 & -11.196152781 -  i 8.196151933 & 13.875544804 \\ \hline
13 & 0.000000084 -  i 15.347547346 & 15.347547346 \\ \hline
14 & 0.000000083 -  i 1.732050808 & 1.732050808 \\ \hline
15 & -101.423006915 -  i 32.954328677 & 106.642459228 \\ \hline
16 & -1.224744868 +  i 1.224744875 & 1.732050808 \\ \hline
17 & -20.645987906 +  i 15.591127116 & 25.871607244 \\ \hline
18 & -40.127945922 -  i 30.808776018 & 50.590836361 \\ \hline
19 & 65.239497521 -  i 83.819706393 & 106.216454548 \\ \hline
20 & -115.811330427 -  i 84.141852139 & 143.150674244 \\ \hline
21 & -53.061367118 -  i 135.569663547 & 145.583798394 \\ \hline
22 & 36.010672960 +  i 16.445523443 & 39.588177633 \\ \hline
23 & -40.070955963 -  i 2.740927872 & 40.164588848 \\ \hline
24 & 164.290883844 -  i 129.064834808 & 208.923972053 \\ \hline
25 & 274.923532195 -  i 34.730921473 & 277.108616721 \\ \hline
26 & 0.000000762 -  i 277.056941014 & 277.056941014 \\ \hline
27 & 152.288761117 +  i 30.720691710 & 155.356453557 \\ \hline
28 & -0.000003253 +  i 136.433611353 & 136.433611353 \\ \hline
29 & -4.270422302 +  i 12.674160517 & 13.374260781 \\ \hline
30 & 485.145409358 +  i 667.745345688 & 825.378649414 \\ \hline
31 & 13.416605095 +  i 0.680413518 & 13.433847358 \\ \hline
32 & 164.960256203 +  i 68.328775084 & 178.551694562 \\ \hline
33 & -82.729230504 +  i 287.059281883 & 298.742626511 \\ \hline
34 & -538.981501170 -  i 268.380890132 & 602.104111256 \\ \hline
35 & -415.659227492 +  i 629.696787632 & 754.513510650 \\ \hline
36 & -681.771583909 +  i 223.335013764 & 717.419696551 \\ \hline
37 & 101.630605413 -  i 150.367371847 & 181.491395038 \\ \hline
38 & -59.482939675 +  i 173.267881065 & 183.193828283 \\ \hline
39 & -279.481438371 -  i 847.784737546 & 892.663898458 \\ \hline
40 & 347.379313920 -  i 1069.123643311 & 1124.143119192 \\ \hline
41 & -941.842709163 -  i 512.157840150 & 1072.088308877 \\ \hline
42 & 399.545316356 -  i 429.453517605 & 586.572061733 \\ \hline
43 & 390.071778264 +  i 279.072428091 & 479.622155784 \\ \hline
44 & 24.267771316 +  i 37.761389387 & 44.887049949 \\ \hline
45 & 2753.539308039 +  i 289.408610795 & 2768.706568944 \\ \hline
46 & 32.920210846 -  i 30.745331135 & 45.044596443 \\ \hline
47 & 536.002877882 -  i 206.437637985 & 574.382784800 \\ \hline
48 & 394.055675294 +  i 817.737030229 & 907.729985094 \\ \hline
49 & -1754.712300120 +  i 400.501606784 & 1799.837990828 \\ \hline
50 & 137.556258644 +  i 2186.393963508 & 2190.716843400 \\ \hline
\end{tabular}
}
\vskip 0.2truecm
\centerline{\bf Table 3}

\pagebreak
\centerline{
\begin{tabular}{|c|c|c|}
\hline
$\bullet$ & $L_{15/2}$ & $\bullet$ \\ \hline
$k$ & $I_k$ & $|I_k|$ \\ \hline
3 & 1.000000000 +  i 0.000000000 & 1.000000000 \\ \hline
4 & 0.000000000 -  i 1.732050808 & 1.732050808 \\ \hline
5 & 0.000000000 +  i 0.000000000 & 0.000000000 \\ \hline
6 & 3.000000000 -  i 3.464101615 & 4.582575695 \\ \hline
7 & 0.000000000 -  i 1.732050808 & 1.732050808 \\ \hline
8 & 1.732050808 +  i 0.000000000 & 1.732050808 \\ \hline
9 & 10.730551990 -  i 5.147276559 & 11.901226911 \\ \hline
10 & 0.000000000 +  i 0.000000000 & 0.000000000 \\ \hline
11 & -12.336706536 +  i 19.196290072 & 22.818673947 \\ \hline
12 & -11.196152423 +  i 8.196152423 & 13.875544804 \\ \hline
13 & 14.350206054 -  i 5.442315293 & 15.347547346 \\ \hline
14 & 0.000000000 +  i 1.732050808 & 1.732050808 \\ \hline
15 & 0.000000000 +  i 0.000000000 & 0.000000000 \\ \hline
16 & 1.224744871 -  i 1.224744871 & 1.732050808 \\ \hline
17 & 9.345902508 +  i 24.124555285 & 25.871607244 \\ \hline
18 & -6.617211192 -  i 50.156208387 & 50.590836361 \\ \hline
19 & 50.553444640 -  i 93.414583721 & 106.216454548 \\ \hline
20 & 0.000000000 +  i 0.000000000 & 0.000000000 \\ \hline
21 & -53.061366041 +  i 135.569663969 & 145.583798394 \\ \hline
22 & -11.153278505 +  i 37.984578277 & 39.588177633 \\ \hline
23 & -29.353726021 -  i 27.414466365 & 40.164588848 \\ \hline
24 & -164.290886665 -  i 129.064831217 & 208.923972053 \\ \hline
25 & 0.000000000 +  i 0.000000000 & 0.000000000 \\ \hline
26 & 228.013392388 +  i 157.386281027 & 277.056941014 \\ \hline
27 & 56.698638283 +  i 144.640561664 & 155.356453557 \\ \hline
28 & 0.000000000 -  i 136.433611353 & 136.433611353 \\ \hline
29 & 11.816320486 +  i 6.264616638 & 13.374260781 \\ \hline
30 & 0.000000000 +  i 0.000000000 & 0.000000000 \\ \hline
31 & -1.359079795 -  i 13.364922631 & 13.433847358 \\ \hline
32 & 164.960256101 +  i 68.328775330 & 178.551694562 \\ \hline
33 & -224.792217910 -  i 196.762841162 & 298.742626511 \\ \hline
34 & -110.636340119 -  i 591.852144574 & 602.104111256 \\ \hline
35 & 0.000000000 +  i 0.000000000 & 0.000000000 \\ \hline
36 & 147.472006710 +  i 702.099015977 & 717.419696551 \\ \hline
37 & 45.731849880 +  i 175.635202563 & 181.491395038 \\ \hline
38 & -177.588145856 -  i 44.971426177 & 183.193828283 \\ \hline
39 & -875.291194999 -  i 175.254556482 & 892.663898458 \\ \hline
40 & 0.000000000 +  i 0.000000000 & 0.000000000 \\ \hline
41 & 1052.478012596 +  i 204.116082250 & 1072.088308877 \\ \hline
42 & 399.545318063 +  i 429.453516017 & 586.572061733 \\ \hline
43 & -171.335849479 -  i 447.974819607 & 479.622155784 \\ \hline
44 & 44.430164502 +  i 6.388093254 & 44.887049949 \\ \hline
45 & 0.000000000 +  i 0.000000000 & 0.000000000 \\ \hline
46 & -17.945816315 -  i 41.315412930 & 45.044596443 \\ \hline
47 & 571.498128245 +  i 57.493242102 & 574.382784800 \\ \hline
48 & -817.737034535 -  i 394.055666358 & 907.729985094 \\ \hline
49 & -780.920437266 -  i 1621.597997004 & 1799.837990828 \\ \hline
50 & 0.000000000 +  i 0.000000000 & 0.000000000 \\ \hline
\end{tabular}
}
\vskip 0.2truecm
\centerline{\bf Table 4}
\pagebreak

\centerline{
\begin{tabular}{|c|c|c|}
\hline
$\bullet$ & $L_{15/4}$ & $\bullet$ \\ \hline
$k$ & $I_k$ & $|I_k|$ \\ \hline
3 & 1.000000000 +  i 0.000000000 & 1.000000000 \\ \hline
4 & 0.000000000 +  i 1.732050808 & 1.732050808 \\ \hline
5 & -1.018073921 - i 3.133309346 & 3.294556414 \\ \hline
6 & 3.000000000 +  i 3.464101615 & 4.582575695 \\ \hline
7 & -0.751508681 -  i 1.560523855 & 1.732050808 \\ \hline
8 & -1.732050808 +  i 0.000000000 & 1.732050808 \\ \hline
9 & 10.730551990 +  i 5.147276559 & 11.901226911 \\ \hline
10 & -15.603243133 -  i 11.336419711 & 19.286669182 \\ \hline
11 & 17.245203123 +  i 14.943053568 & 22.818673947 \\ \hline
12 & -11.196152423 -  i 8.196152423 & 13.875544804 \\ \hline
13 & 3.672908488 +  i 14.901575513 & 15.347547346 \\ \hline
14 & -1.688624678 +  i 0.385417563 & 1.732050808 \\ \hline
15 & 0.000000000 +  i 0.000000000 & 0.000000000 \\ \hline
16 & -1.224744871 +  i 1.224744871 & 1.732050808 \\ \hline
17 & -17.429589094 -  i 19.119348456 & 25.871607244 \\ \hline
18 & -6.617211192 +  i 50.156208387 & 50.590836361 \\ \hline
19 & 0.000000000 -  i 106.216454548 & 106.216454548 \\ \hline
20 & -44.235991098 +  i 136.144381552 & 143.150674244 \\ \hline
21 & 72.909410760 -  i 126.011349400 & 145.583798394 \\ \hline
22 & -36.010673013 +  i 16.445523326 & 39.588177633 \\ \hline
23 & -10.836276386 +  i 38.675176941 & 40.164588848 \\ \hline
24 & 164.290886665 -  i 129.064831217 & 208.923972053 \\ \hline
25 & -257.649075864 +  i 102.010485575 & 277.108616721 \\ \hline
26 & 275.036883975 -  i 33.395523912 & 277.056941014 \\ \hline
27 & -153.611719961 +  i 23.217819719 & 155.356453557 \\ \hline
28 & 106.668092622 +  i 85.064965309 & 136.433611353 \\ \hline
29 & -9.197471902 +  i 9.709653035 & 13.374260781 \\ \hline
30 & 0.000000000 +  i 0.000000000 & 0.000000000 \\ \hline
31 & -4.021598499 +  i 12.817761129 & 13.433847358 \\ \hline
32 & -164.960256101 -  i 68.328775330 & 178.551694562 \\ \hline
33 & 85.599713692 +  i 286.216431937 & 298.742626511 \\ \hline
34 & -217.505092368 -  i 561.445362956 & 602.104111256 \\ \hline
35 & 33.851120653 +  i 753.753765751 & 754.513510650 \\ \hline
36 & 147.472006710 -  i 702.099015977 & 717.419696551 \\ \hline
37 & -136.240563896 +  i 119.906777214 & 181.491395038 \\ \hline
38 & 0.000000000 +  i 183.193828283 & 183.193828283 \\ \hline
39 & 538.949599873 -  i 711.605343156 & 892.663898458 \\ \hline
40 & -909.450887536 +  i 660.754746927 & 1124.143119192 \\ \hline
41 & 1008.985242455 -  i 362.383943545 & 1072.088308877 \\ \hline
42 & -546.310790198 +  i 213.568031595 & 586.572061733 \\ \hline
43 & 426.548198677 +  i 219.303548819 & 479.622155784 \\ \hline
44 & -24.267771377 +  i 37.761389348 & 44.887049949 \\ \hline
45 & 0.000000000 +  i 0.000000000 & 0.000000000 \\ \hline
46 & -6.133571758 +  i 44.625048641 & 45.044596443 \\ \hline
47 & -558.735830232 -  i 133.153503483 & 574.382784800 \\ \hline
48 & 394.055666358 +  i 817.737034536 & 907.729985094 \\ \hline
49 & -883.212092863 -  i 1568.232505801 & 1799.837990828 \\ \hline
50 & 410.499402001 +  i 2151.913225229 & 2190.716843400 \\ \hline
\end{tabular}
}
\vskip 0.2truecm
\centerline{\bf Table 5}
\end{document}